# Excitonic interactions and mechanism for ultrafast interlayer photoexcited response in van der Waals heterostructures

Chen Hu[1,2], Mit H. Naik[1,2], Yang-Hao Chan[1,2,3], and Steven G. Louie[1,2*]

*[1] Department of Physics, University of California at Berkeley, Berkeley, California 94720, USA*

*[2] Materials Sciences Division, Lawrence Berkeley National Laboratory, Berkeley, California 94720, USA*

*[3] Institute of Atomic and Molecular Sciences, Academia Sinica, and Physics Division, National Center for Theoretical Sciences, Taiwan*

*Email: sglouie@berkeley.edu

## Abstract

Optical dynamics in van der Waals heterobilayers is of fundamental scientific and practical interest. Based on a time-dependent adiabatic *GW* approach, we discover a new many-electron (excitonic) channel for converting photoexcited intralayer to interlayer excitations and the associated ultrafast optical responses in heterobilayers, which is conceptually different from the conventional single-particle picture. We find strong electron-hole interactions drive the dynamics and enhance the pump-probe optical responses by an order of magnitude with a rise time of ~300 fs in $MoSe_2/WSe_2$ heterobilayers, in agreement with experiment.

By stacking a variety of two-dimensional (2D) material layers together, van der Waals (vdW) heterostructures offer a novel, versatile platform for exploring fascinating low-dimensional physics, such as Hofstadter butterfly states [1], topological phases [2,3], unconventional superconductors [4], and strong-correlated insulators [5]. Among the diverse 2D materials, the semiconducting monolayer transition metal dichalcogenide (TMD) family has attracted broad interest for photonics and optoelectronics because of their direct band gap properties and strong light-matter interactions [6]. Particularly, many TMD heterobilayers exhibit a type-II band alignment, *i.e.* the conduction band minimum (CBM) and the valence band maximum (VBM) reside in different layers. The type-II alignment provides a rich playground for investigating various optically-induced exciton states (intralayer excitons on the individual layers and interlayer excitons across the two layers), as well as their dynamical evolution from mutual interactions [7-19].

Although recent optical pump-probe spectroscopies have provided rich experimental signatures on the ultrafast optical responses in heterobilayers [7-10], their accurate theoretical description and understanding remain to be fully understood. Most previous experimental interpretations on the interlayer optical dynamics are based on a "single-particle" picture: independent holes or electrons transfer from one monolayer to its neighbor through band hybridization or phonon-assisted scattering [7]. However, such an independent-particle (IP) picture is conceptually insufficient: in atomically thin TMD materials, the electrons and holes are known to be strongly bounded to form two-particle low energy states, *i.e.* excitons (correlated electron-hole pairs) with giant binding energy, due to enhanced Coulomb interaction from low-dimensional quantum confinement and reduced dielectric screening [6,20]. A vast literature of previous studies has shown that the IP picture can not describe most optically-induced phenomena even qualitatively in low-dimensional systems [20-26]. Therefore, a many-electron picture including electron-hole interactions (excitonic effects) are important to establish a more comprehensive interpretation of photoexcitation dynamics in TMD heterobilayers. However, this is significantly challenging in *ab initio* calculations because both time-dependent nonequilibrium dynamics and many-body excitonic physics need be accurately captured simultaneously.

In this study, going beyond the IP picture, we discover a *many-body excitonic mechanism*, as an alternative or additional channel, for the dynamics of converting photoexcited intralayer to interlayer excitations and for the associated ultrafast optical responses in TMD heterobilayers. We find that the couplings between the intralayer and interlayer excitonic state induce an ultrafast optical response, conceptually different from the direct single-particle charge transfer channel in the IP picture. Through our *ab initio* time-dependent calculations, strong excitonic effects are shown to play a crucial role on the ultrafast exciton dynamics and the pump-probe optical responses by enhancing the signal by over one-order of magnitude with a rise time about of 300 femtoseconds. To investigate the real-time nonequilibrium dynamics including many-electron interactions and excitonic effects from first-principles, we employ a recently developed *ab initio* time-dependent adiabatic $GW$ approach (TD-a$GW$) [26] with real-time propagation of the density matrix in the presence of an external optical field.

In the TD-a$GW$ framework [26],

$$i\hbar \frac{\partial}{\partial t}\rho_{nm,\mathbf{k}}(t) = [H^{aGW}(t), \rho(t)]_{nm,\mathbf{k}}, \qquad (1)$$

where $n$ and $m$ are band indices, and $\rho_{nm,\mathbf{k}}(t)$ is the density matrix of the interacting many-electron system in the Bloch-state basis which serves as the key quantity to explore the time-dependent field-driven nonequilibrium system. $H^{aGW}_{nm,\mathbf{k}}(t)$ is the TD-a$GW$ Hamiltonian defined as:

$$H^{aGW}_{nm,\mathbf{k}}(t) = h_{nm,\mathbf{k}} + U^{ext}_{nm,\mathbf{k}}(t) + \Delta V^{ee}_{nm,\mathbf{k}}(t). \qquad (2)$$

Here, $h_{nm,\mathbf{k}}$ is the equilibrium quasi-particle Hamiltonian which includes all the equilibrium interactions (before photoexcitation) at the $GW$ level, and it is independent of time. The external field part is given by $U^{ext}_{nm,\mathbf{k}}(t)$ which denotes the light-matter interaction and is equal to $-e\mathbf{E}(t)\cdot \mathbf{d}_{nm,\mathbf{k}}$, where $\mathbf{E}(t)$ is the optical electric field and $\mathbf{d}_{nm,\mathbf{k}}$ is the optical dipole matrix. Importantly, the photo-induced variation of the electron-electron interaction is given by $\Delta V^{ee}_{nm,\mathbf{k}}(t)$ which has two parts: $\Delta V^{ee}_{nm,\mathbf{k}}(t) = \Delta V^{H}_{nm,\mathbf{k}}(t) + \Delta \Sigma^{COHSEX}_{nm,\mathbf{k}}(t)$, where the first term is the Hartree potential and the second term is the nonlocal Coulomb hole plus screened-exchange (COHSEX) $GW$ self-energy taken in the static screening limit which accounts for the time change of the many-electron (quasiparticle and excitonic) effects. In the TD-a$GW$ approach, the excitonic effects is accurately captured at the $GW$ plus Bethe-Salpeter equation ($GW$-BSE) level [26]. This is validated by the identical linear absorption spectra (Fig. S2 in the Supplemental Material [27]) comparing results from standard $GW$-BSE calculations with those from TD-a$GW$. During the real-time evolution, we focus on the photoinduced time variation of the various quantities: $\Delta\rho(t) = \rho(t) - \rho(eq.)$, $\Delta V^{ee}(t) = V^{ee}(t) - V^{ee}(eq.)$, etc., where $eq.$ denote the equilibrium quantities (i.e., in the absence of a time-dependent driving external field). The formalism and computational details on the *ab initio* TD-a$GW$ method can be found in the Supplemental Material [27].

Figure 1(a-b) illustrate the theoretical pump-probe setup. We focus on a representative and well studied TMD heterobilayer - MoSe$_2$/WSe$_2$ with a common unit cell size of lattice constant 3.30 Å.

As shown in Fig. 1(c), its quasiparticle band structure exhibits a type-II band alignment at the $K$ point (as well as the $K'$ point): the CBM (VBM) is mainly from the MoSe$_2$ (WSe$_2$) layer with a quasiparticle bandgap of 1.77 eV. There is very little layer hybridization for states near the $K$ point, consistent with previous work [25]. Fig. 2(a) shows the equilibrium linear optical absorbance of the system, which is computed by solving the $GW$-BSE (computational details are given in Supplemental Material [27]). Without electron-hole interactions, the IP picture fails to capture the main features seen in experiments (not shown) even qualitatively. On the other hand, the results with excitonic effects (red curve) show prominent absorption peaks corresponding to: intralayer MoSe$_2$ excitons (electron and hole reside on the MoSe$_2$ layer), intralayer WSe$_2$ excitons (electron and hole reside on the WSe$_2$ layer), as well as interlayer excitons (electron and hole reside on the MoSe$_2$ and WSe$_2$, respectively) which have lower energies but weak optical response due to their small optical oscillator strengths.

Next, we investigate the real-time evolution of the interlayer and intralayer exciton states after the system is optically pumped at the MoSe$_2$ 1s exciton excitation energy (1.79 eV, the first peak in Fig. 2(a)), as shown in Fig. 1(b). In the IP picture, the main dynamics of the system would be that of independent carriers (e.g., photo-induced holes transferring from MoSe$_2$ layer to WSe$_2$ layer) through band hybridization or phonon-assisted scatterings. However, for 2D materials, the photoexcited holes are physically not "independent" but bound to the excited electrons forming correlated electron-hole pairs (excitons) with large binding energies (> 300 meV). Therefore, as another channel for the dynamics going beyond the IP picture, we propose a new many-body excitonic picture as shown in Fig. 2(b): after pumping, the excited 1s exciton of MoSe$_2$ couples to the corelated-continuum (CC) states of the interlayer excited states which are in resonance in energy. Our TD-a$GW$ calculations reveal that the dynamical coupling between the photoexcited MoSe$_2$ intralayer 1s exciton state and the interlayer excitonic continuum states in fact gives rise to a fast and significant time evolution in the system.

The computed real-time dynamics of an averaged photoinduced electron-hole coherence are shown in Fig. 2(c-d). Here we define the averaged electron-hole coherence between two bands $\langle \Delta \rho_{vc}(t) \rangle$ from the change in the off-diagonals of the density matrix as: $\langle \Delta \rho_{vc}(t) \rangle = \frac{1}{N_k} \sum_k |\Delta \rho_{vc,k}(t)|^2$. This quantity provides a measure of the correlation between the electrons and holes in the conduction band $c$ and valence band $v$, respectively. As discussed below, the individual k-dependent $\Delta \rho_{vc,k}(t)$ itself is of course the key quantity for determining the measured time-dependent optical responses. As shown in Fig. 2(c), after resonantly pumping at the MoSe$_2$ 1s exciton excitation energy (1.79 eV), the pump-induced $\langle \Delta \rho_{vc}(t) \rangle$ demonstrates two significant evolution dynamics: (1) During the pump-pulse duration (marked by the grey shadow in Fig. 2(c), with a full-width-at-half-maximum of 100 fs), the intralayer MoSe$_2$ exciton is excited. The magnitude of $\Delta\rho_{v_{Mo}c_{Mo}}$ increases dramatically and reaches a maximum at the end of the pump pulse (at about 120 fs). In the upper left panel of Fig. 2(d), at $t = 120$ fs, a $k$-space distribution analysis reveals that the photoexcited intralayer $\Delta\rho_{v_{Mo}c_{Mo}}(\mathbf{k})$ is centered at $K$ point, showing typical 1s exciton features. The lower left panel of Fig. 2(d) indicates a very small interlayer $\Delta\rho_{v_Wc_{Mo}}(\mathbf{k})$. (2) After the pump pulse ended (>120 fs), the magnitude of $\Delta\rho_{v_{Mo}c_{Mo}}$ shrinks with time along with a growth of the interlayer $\Delta\rho_{v_Wc_{Mo}}$, demonstrating the excited system is evolving rapidly from one of intralayer exciton character to one with increasing interlayer exciton character. Interestingly, as shown in the lower right panel of Fig. 2(d), the interlayer $\Delta\rho_{v_Wc_{Mo}}(\mathbf{k})$ is not concentrated at $K$ point, reflecting a resonant coupling between the intralayer (MoSe$_2$) 1s exciton and interlayer correlated-continuum excited states as schematically indicated in Fig. 2(b).

We note that the coherence between holes in different layers ($\Delta\rho_{v_{Mo}v_W}$) is significantly small, therefore a direct valence-to-valence band coupling is negligible, verifying that the single-particle picture of hole transition between valence bands of the two layers is not significant in the ultrafast time scale (Fig. S3 of the Supplemental Material [27]). Very fast phonon-assisted processes are needed if such mechanism is to be viable in the time regime of a few tens to hundreds of femtoseconds.

We now will discuss the driving forces that give rise to the dynamics. In the TD-a$GW$ Hamiltonian (Eq. (2)), there are two time-dependent terms: the external field $U^{ext}(t)$ and the internal electron-electron interaction $\Delta V^{ee}(t)$. The latter can be further spitted into two parts: $\Delta V^H(t)$ and $\Delta\Sigma^{COHSEX}(t)$, as defined above. In Fig. 3, we compare the real-time evolution of the expectation values of these dynamical terms. Our *ab initio* results show that the $\Delta\Sigma^{COHSEX}(t)$ term which accounts for the excitonic effects is dominant and much larger than the external field term and the Hartree term: the maximum magnitude ratios are $\langle\Delta\Sigma^{COHSEX}\rangle/\langle\Delta V^H\rangle \approx 10$ and $\langle\Delta\Sigma^{COHSEX}\rangle/\langle U^{ext}\rangle \approx 30$, revealing the crucial role of strong excitonic effects. Therefore, excitonic effects is the dominant driving force and must be accurately included in the real-time evolution.

In experiments, the ultrafast pump-induced intralayer-to-interlayer dynamical evolution of the system is measured by an optical pump-probe setup [8] as shown schematically in Fig. 1(b): a much weaker optical pulse is employed to probe one layer (e.g., WSe$_2$) after a time delay $\tau$ from a strong pump pulse on the other layer (e.g., MoSe$_2$), and the transient probe response signal is used to investigate the real-time system dynamics. Optically probing the interlayer excitation might offer another way to track the dynamics of interlayer excited states, but typically the very small oscillator strength of the interlayer transitions makes it experimentally impractical (as seen in the extremely weak computed interlayer optical response in Fig. 2(a)). Instead, the WSe$_2$ 1s exciton (1.90 eV, the second peak in Fig. 2(a)) is optically probed in previous experiments after a pump pulse is used to excite the MoSe$_2$ 1s exciton (the first peak in Fig. 2(a)). A widely-measured quantity is the transient absorption change $\Delta A/A$ defined as $[A_{pump\ on}(\tau) - A_{pump\ off}]/A_{pump\ off}$, where $A_{pump\ on}$ and $A_{pump\ off}$ are absorptions at the probed energy after delay time $\tau$, with and without the pump field [7-11], respectively. In our theoretical study, the transient optical properties are calculated from the polarization $\boldsymbol{P}(t) = \frac{e}{N_k V}\sum_{nm,\boldsymbol{k}} \rho_{nm,\boldsymbol{k}}(t)\, \boldsymbol{d}_{mn,\boldsymbol{k}}$, where $N_k$ is the number of $k$ points in the BZ sampling and $V$ is the volume of the unit cell. From this expression, it is clear that the optical dynamics is intrinsically related to the dynamics of $\rho_{nm,\boldsymbol{k}}(t)$ discussed above. More details on the calculation of the transient optical absorption from our *ab initio* TD-a$GW$ method are given in the Supplemental Material [27].

As shown in Fig. 4, the computed transient absorption change $\Delta A/A$ (in percents) from TD-a$GW$ including excitonic effects exhibits a fast and prominent rise during and immediately after the pump duration (grey shadow in the plot). By fitting to an exponential form $c(1 - e^{-\tau/t_r})$ we extract a theoretical rise time of $t_r \approx 300$ fs, which agrees well with the experimental measurements of about 200 fs [8,13]. On the contrary, in the case of TD-Hartree calculation in which only the Hartree term $\Delta V^H(t)$ is included but ignoring excitonic effects from $\Delta \Sigma^{COHSEX}(t)$, there is negligible change in the optical response with the pump pulse: the magnitude is about 30 times smaller than the full TD-a$GW$ case and no discernable rising behavior exists after the pump pulse is subsided, which qualitatively disagree with experiments [8,13]. These results further evidence the key role of excitonic effects on the optical dynamics, which is consistent with the driving forces analysis in Fig. 3. In experiments, existing literature shows that difficult-to-control measurement deviations happen due to variations in relative layer orientation, sample quality, pump-probe details, and other technical issues, leading to a range for the value of the rise time from several tens of fs to hundreds of fs in various TMD heterobilayer samples [7-13]. Moreover, the effects of exciton-phonon coupling which lies outside of the scope of this paper, is expected to bring additional scattering channels into the time evolution [14-19]. As a result, it may also contribute to the rise time which calls for future studies.

In summary, we discover a new many-body excitonic channel for the ultrafast dynamics of photoinduced excited states and optical responses in TMD heterobilayers. Employing a recently developed *ab initio* TD-a*GW* method, we investigate the real-time evolution of the inter-/intra-layer electron-hole coherence, and demonstrate that the couplings between the intralayer 1s exciton states and the interlayer correlated-continuum excited states in a $MoSe_2$/$WSe_2$ heterobilayer give rise to real-time excited-state and optical dynamics that are consistent with experiments. The strong excitonic effects in the van der Waals heterobilayers are shown as a major driving force for their excited-state dynamics and ultrafast pump-probe optical responses. The many-body excitonic picture, going beyond the independent-particle theory, thus provides new insights to the ultrafast optoelectronics in TMD heterostructures.

**Acknowledgments** This work was supported by Center for Computational Study of Excited-State Phenomena in Energy Materials (C2SEPEM) at LBNL, funded by the U.S. Department of Energy, Office of Science, Basic Energy Sciences, Materials Sciences and Engineering Division under Contract No. DE-AC02-05CH11231, as part of the Computational Materials Sciences Program which provided advanced codes and simulations, and supported by the Theory of Materials Program (KC2301) funded by the U.S. Department of Energy, Office of Science, Basic Energy Sciences, Materials Sciences and Engineering Division under Contract No. DE-AC02-05CH11231 which provided conceptual and theoretical developments of exciton couplings. Y.-H. C. is supported by the National Science and Technology Council of Taiwan under grant no. 110-2124-M-002-012. Computational resources were provided by National Energy Research Scientific Computing Center (NERSC), which is supported by the Office of Science of the US Department of Energy under contract no. DE-AC02-05CH11231, Stampede2 at the Texas Advanced Computing Center (TACC), The University of Texas at Austin through Extreme Science and Engineering Discovery Environment (XSEDE), which is supported by National Science Foundation under grant no. ACI-1053575 and Frontera at TACC, which is supported by the National Science Foundation under grant no. OAC-1818253. C. H. thanks Y. Yoon for fruitful discussions on pump-probe experiments, and Z. Li, M. Wu, J. Ruan, and J. Jiang for useful help on many-body theories and computations.

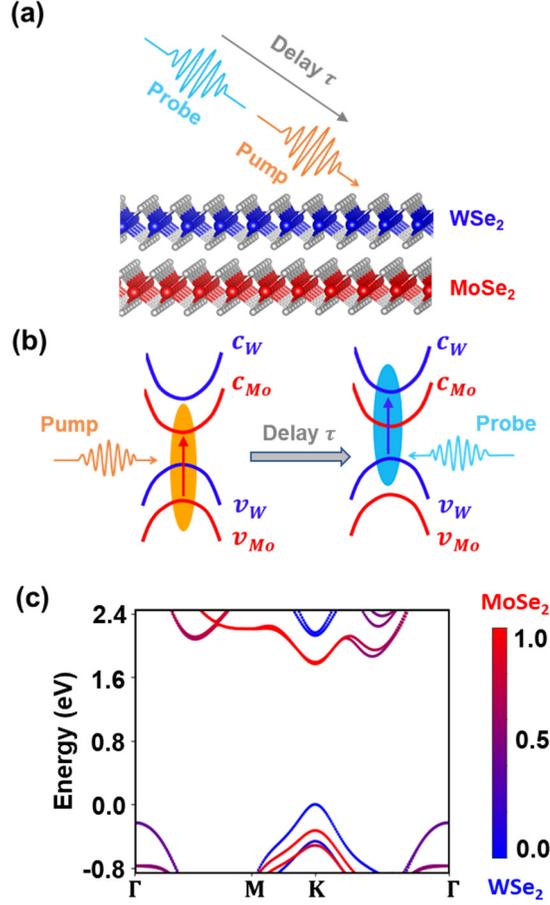

**Fig. 1**: **(a-b)** Schematic of the ultrafast optical pump-probe setup for study of MoSe$_2$/WSe$_2$ heterobilayers with a type II band alignment. The delay time $\tau$ denotes the time separation between the pump pulse and the probe pulse. Optically-pumped excitation of 1s exciton in the MoSe$_2$ layer leads to the ultrafast inter-layer dynamics which is probed by an optical pulse with WSe$_2$ 1s exciton excitation energy after different time delay $\tau$. $c_W, c_{Mo}, v_W, v_{Mo}$ denote the conduction band of WSe$_2$ and MoSe$_2$, and the valence band of WSe$_2$ and MoSe$_2$, respectively. **(c)** $GW$ quasiparticle band structure. The energy of the top of valence band of the bilayer is set to be zero. The color represents the layer-component projection of each Bloch state to the different layers.

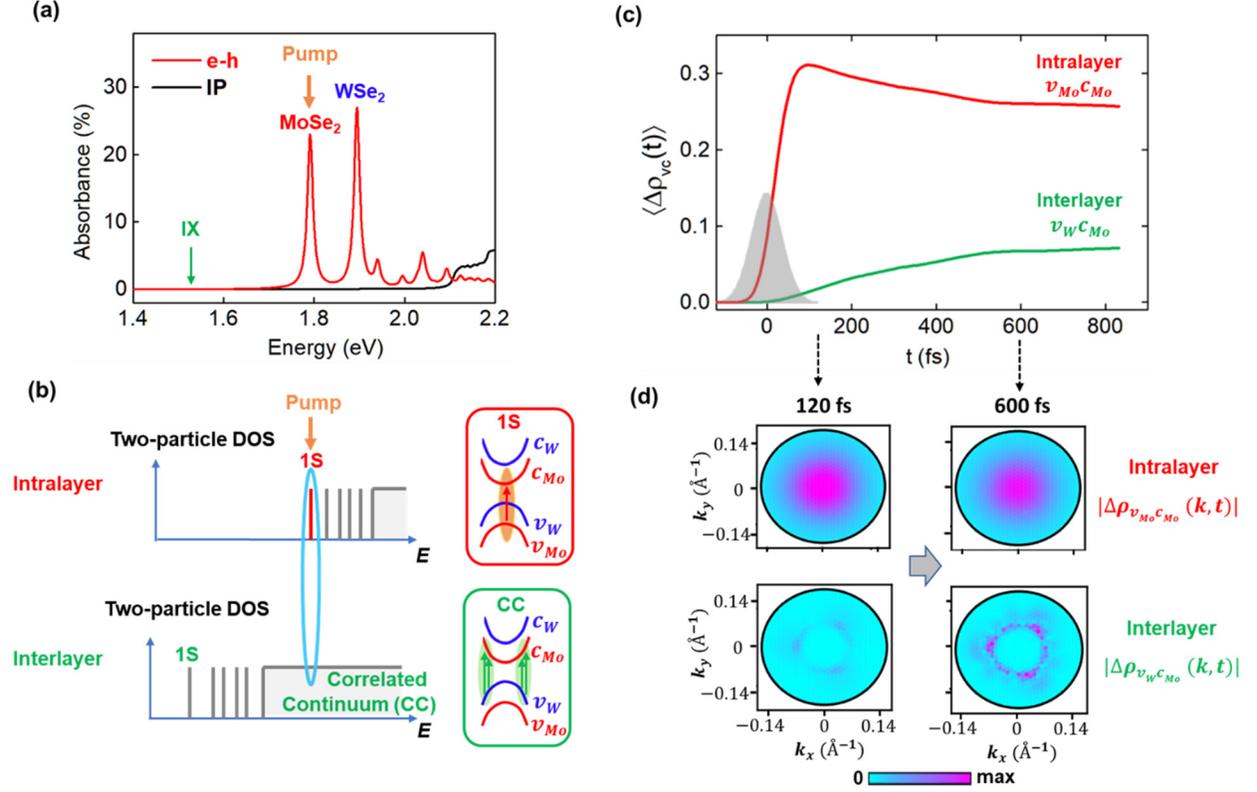

**Fig. 2**: **(a)** Absorbance of the MoSe$_2$/WSe$_2$ heterobilayer computed with ($e$-$h$, red solid line) and without (IP, black dashed line) electron-hole interactions. The position of the lowest-energy optically active interlayer exciton (IX) is marked by the green arrow. **(b)** Schematic of coupling between intralayer (MoSe$_2$) 1s exciton and interlayer correlated-continuum (CC) excited states, indicated by the two-particle (correlated electron-hole pair) density of states (DOS). **(c)** Pump-induced real-time evolution of electron-hole coherence, as seen through an average of the photoinduced variations of the density matrix off-diagonals: $\langle \Delta\rho_{vc}(t) \rangle = \frac{1}{N_k}\sum_{k}|\Delta\rho_{vc,k}(t)|^2$, where intralayer (red line) and interlayer (green line) quantities are given by taking $vc = v_{Mo}c_{Mo}$ and $v_W c_{Mo}$ respectively. The grey shadow area shows the pump-pulse intensity duration in time (Gaussian shape). We set the center of the pump pulse as zero of the time line. **(d)** $k$-space distributions of intralayer and interlayer electron-hole coherence magnitudes $|\Delta\rho_{vc}(\boldsymbol{k},t)|$ at two representative time points: snapshot at 120 femtoseconds (fs) and at 600 femtoseconds (fs) after the pump pulse, showing significant interlayer dynamics. In the plot, the **K** point is set as the origin for clarity.

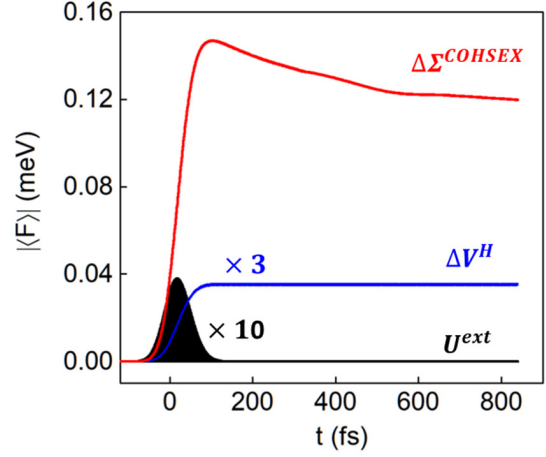

**Fig. 3**: Real-time evolution of various terms in the TD-a$GW$ effective Hamiltonian. The expectation values of time dependence of different dynamical terms are defined as $\langle F(t)\rangle = Tr\{\rho(t)F(t)\}$, where $F$ is the external optical pump field ($U^{ext}$, black line), pump-induced Hartree potential variation ($\Delta V^H$, blue line), or pump-induced many-body COHSEX self-energy variation ($\Delta\Sigma^{COHSEX}$, red line).

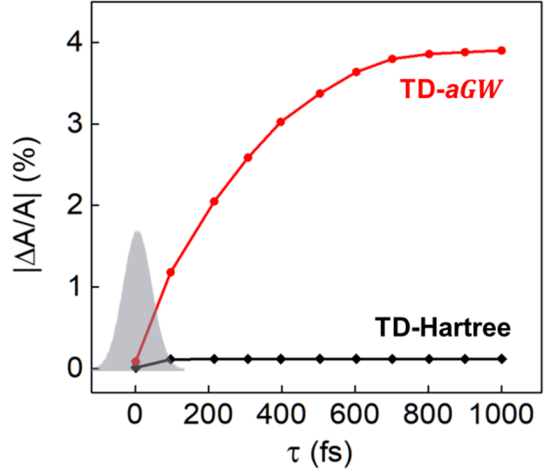

**Fig. 4**: Computed transient optical response as function of pump-probe delay time $\tau$, as schematically depicted in Fig. 1(b). The black and red lines present the TD-a$GW$ and TD-Hartree results, which are time evolutions with and without excitonic effects (*i.e.,* many-body $\Delta\Sigma^{COHSEX}(t)$ term) included, respectively. The grey shadow area shows the pump-pulse intensity duration in time (Gaussian shape).

# Supplemental Material for "Excitonic interactions and mechanism for ultrafast interlayer photoexcited response in van der Waals heterostructures"

Chen Hu[1,2], Mit H. Naik[1,2], Yang-Hao Chan[1,2,3], and Steven G. Louie[1,2*]

*[1] Department of Physics, University of California at Berkeley, Berkeley, California 94720, USA*

*[2] Materials Sciences Division, Lawrence Berkeley National Laboratory, Berkeley, California 94720, USA*

*[3] Institute of Atomic and Molecular Sciences, Academia Sinica, and Physics Division, National Center for Theoretical Sciences, Taiwan*

\*Email: sglouie@berkeley.edu

In this Supplemental Material, we provide some detailed computational information and further results. We organize the material into the following sections:

1. Computational details of DFT calculations, $GW$-BSE calculations, and TD-a$GW$ calculations;
2. Geometric configuration of the $MoSe_2$/$WSe_2$ heterobilayer;
3. Comparison of optical absorbances computed from TD-a$GW$ and $GW$-BSE;
4. Hole-hole coherence dynamics.

# I. Computational details

The workflow of this work involves three main computation steps: **(1)** Calculations of ground-state properties based on density functional theory (DFT) with the Quantum Espresso package [28]; **(2)** Calculations of equilibrium quasiparticle states based on the GW method [29] and of excitons and optical properties based on the $GW$ plus Bethe-Salpeter equation ($GW$-BSE) approach [30,31] with the Berkeley$GW$ package [32]; **(3)** Calculations of real-time nonequilibrium excited-state dynamics and transient optical response based on the time-dependent adiabatic $GW$ (TD-a$GW$) approach [26].

**(A) DFT calculation**

In this work, the mean-field DFT calculations are performed with the Quantum Espresso package [28]. We use Optimized Norm-Conserving Vanderbilt (ONCV) pseudopotentials [33] with PBE to the exchange-correlation functional [34], and a plane-wave basis with an energy cutoff of 80 Ry. The structure is fully optimized by minimizing the forces on the atoms with van der Waals (vdW) interactions. The optimized lattice constant is 3.30 Å and the interlayer distance (metal atom to metal atom) is 6.50 Å. The optimized atomic structure is shown in Fig. S1. The full spinor formalism with spin-orbital coupling is employed in the DFT calculations.

**(B) $GW$-BSE calculation**

The $GW$ and $GW$-BSE calculations are performed with the Berkeley$GW$ package [29-32]. For the $GW$ calculations (at the $G_0W_0$ level) on the quasiparticle energies, a total of 6000 bands and a 35 Ry cutoff are included. We use a $6 \times 6 \times 1$ **k**-grid with 10 subsampling points within the nonuniform neck subsampling (NNS) method to describe the 2D dielectric screening [35]. The dynamical screening is treated by using the Hybertsen-Louie generalized plasmon pole model [29].

For the $GW$-BSE calculations on excitons and optical absorption, a very fine **k**-point sampling of $210 \times 210 \times 1$ is taken to well capture the properties of the correlated-continuum. A full interacting kernel is employed both in the BSE and TD-a$GW$ calculations. The spin-orbital coupling of TMD monolayers or heterobilayers is known to give rise to two split exciton series (A and B) with large energy separations (>200 meV in the $MoSe_2$/$WSe_2$ system we studied) and negligible mixing between them [20]. As a result, despite that we include spin-orbital coupling in all calculations (DFT, $GW$-BSE, TD-a$GW$), only the lower-energy A series is focused in our discussions and analysis (consistent with experimental studies [8]). In the optical absorption plot [Fig. 2(a) of the main text], a Lorentzian broadening factor of 7 meV is used.

### (C) TD-a$GW$ calculation

A general approach to solve interacting quantum many-body systems in nonequilibrium is the Kadanoff-Baym equations (KBE) approach, which describes the equations of motion for the interacting Green's function $G$ on the Keldysh contour $C$ [36]:

$$i\frac{d}{dt}G(t,t') = \delta(t,t') + [H_0 + U^{ext}(t)]G(t,t') + \int_C \Sigma(t,t'')G(t'',t')dt'', \quad (S1)$$

where $H_0$ is the static equilibrium Hamiltonian, $U^{ext}(t)$ denotes the interaction with an external field, $G$ is the contour-ordered Green's function, and $\Sigma$ is the electron self-energy operator. A complete set of KBE should combine Eq. S1 with another equivalent adjoint equation of propagating $t'$. Direct solutions for the KBE need the treatment of two coupled time variables $t$ and $t'$, which gives rise to tremendous computational burden for time propagating real material systems to a measurable duration (*e.g.,* several hundreds of fs in ultrafast experiments). A practical and effective approach we adopted is the adiabatic $GW$ (a$GW$) approximation [26,37], where an instantaneous self-energy (time-diagonal $t = t'$), *i.e.,* in the static Coulomb hole plus screened-exchange (COHSEX) [32] limited, is employed.

Within the TD-a$GW$ approach, the KBE in Eq. S1 can be rigorously expressed as the equation of motion of the density matrix in Eq. 1 of the main text [26,37]. In our calculations, a Bloch orbital basis is used: $O_{nm,k} = \langle n\mathbf{k}|\hat{O}|m\mathbf{k}\rangle$, where $|n\mathbf{k}\rangle$ and $|m\mathbf{k}\rangle$ are Bloch eigenstates, and $\hat{O}$ is the physical quantity operator of interest, for example, the density matrix $\rho$, self-energy $\Sigma$, *etc*. As shown in the TD-a$GW$ formalism (Eq. 1 of the main text), at equilibrium, all interactions and information are included in the $h_{nm,k}$ term (the equilibrium quasi-particle energies which are independent of time); therefore, during the time evolution, we only need to focus on the field-induced variation of the quantities: $\Delta\rho_{nm,k}(t) = \rho_{nm,k}(t) - \rho_{nm,k}(eq.)$, $\Delta V^{ee}_{nm,k}(t) = V^{ee}_{nm,k}(t) - V^{ee}_{nm,k}(eq.)$, *etc.,* where *eq.* denotes the equilibrium case. In the Bloch state basis of the TD-a$GW$ approach, excitonic effects are included through $\Delta\Sigma^{COHSEX}_{nm,k}(t)$ as:

$$\Delta\Sigma^{COHSEX}_{nm,k}(t) = i\sum_{n'm',k'} \Delta G^<_{n'm',k'}(t)\, W_{m\mathbf{k},m'\mathbf{k}';n\mathbf{k},n'\mathbf{k}'}, \quad (S2)$$

where $G^<_{n'm',k'}(t)$ is the time-diagonal of the lesser Green's function and we have $\rho_{nm,k}(t) = -iG^<_{nm,k}(t)$. $W$ is the statically screened Coulomb potential as used in standard formulation of the interacting kernel of the BSE Hamiltonian. It is proved that by employing the $\Delta\Sigma^{COHSEX}_{nm,k}(t)$ with TD-a$GW$, one can capture the many-body excitonic effects (electron-hole interaction) in linear optical processes at the $GW$-BSE level [26,37], which is evidenced by the identical linear absorption spectra from the TD-a$GW$ method and $GW$-BSE method as shown in Fig. S2.

The optical properties in the TD-a$GW$ framework are calculated from the optical-induced (probe-induced) polarization variation:

$$\mathbf{\Delta P}(t) = \frac{e}{N_k V} \sum_{nm,\mathbf{k}} \Delta \rho_{nm,\mathbf{k}}(t) \, \mathbf{d}_{mn,\mathbf{k}} , \tag{S3}$$

where $N_k$ is the number of $k$ points, $V$ is the volume of the unit cell, and $\mathbf{d}_{mn,\mathbf{k}}$ is the optical dipole matrix. The linear optical absorption $A(\omega)$ is intrinsically related to the imaginary part of first-order optical susceptibility $Im[\chi(\omega)]$ by: $A(\omega) = 1 - e^{-A_0(\omega)}$ and $A_0(\omega) = \omega d Im[\chi(\omega)]/c$, where $d$ is the thickness of the 2D supercell and $c$ is the speed of light in vacuum [38,39]. The first-order optical susceptibility $\chi(\omega)$ can be calculated from the Fourier transformations of the optical field $\mathbf{E}(t)$ and its induced polarization $\mathbf{\Delta P}(t)$ as:

$$\mathbf{\Delta P}(\omega) = \varepsilon_0 \chi(\omega) \mathbf{E}(\omega), \tag{S4}$$

where $\varepsilon_0$ is the permittivity of vacuum. The probe-induced transient optical response is computed from the probe-induced variation of the polarization of the system, which depends on the delay time $\tau$ of the probe pulse. We follow the scheme described in Refs. [40-42].

In our TD-a$GW$ calculations, the time propagation is performed with a time step of 0.0048 fs using a fourth order Runge-Kutta method. Two valence band and two conduction band ($c_{Mo}, c_W, v_{Mo}, v_W$) are included in the simulations. We employ a very fine $\mathbf{k}$ sampling $210 \times 210 \times 1$ of the Brillouin zone to well capture the properties of the correlated-continuum (consistent with the BSE interacting kernel). The intensity of the pump field is taken as $1.0 \times 10^7 \, W/cm^2$, to excite a total carrier (electron or hole) density of about $5 \times 10^{11} cm^{-2}$, consistent with the order of magnitude in usual experimental setups [9,12]. The intensity of the probe field is taken as $10^{-5}$ that of the pump field so it has negligible effect on the pump-induced dynamics. The pump-pulse field is taken as a Gaussian shape in time with the full width at half-maximum (FWHM) of 100 fs.

## II. Geometric configuration

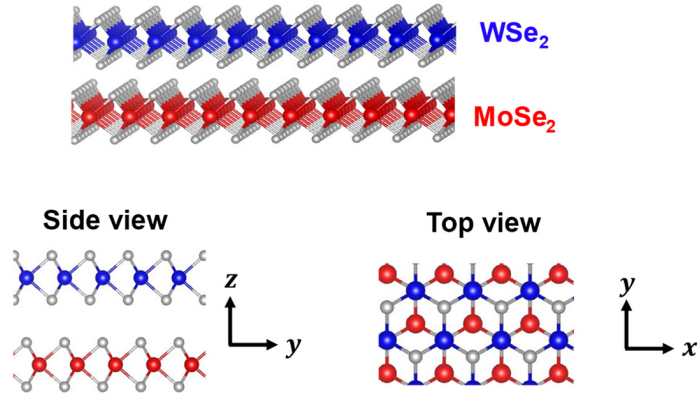

Fig. S1: Atomic structure of the optimized MoSe$_2$/WSe$_2$ bilayer.

The atomic structure (energy-favorable stacking configuration) of MoSe$_2$/WSe$_2$ is fully optimized by minimizing the forces on atoms with vdW interactions. MoSe$_2$ and WSe$_2$ have negligible lattice constant mismatch (<0.1%). Thus, in this study, the lattice constants of the two layers are taken to be the same and relaxed. The purple, light blue and green balls represent Mo, W and Se atoms, respectively. The optimized lattice constant is 3.30 Å and the interlayer distance (metal atom to metal atom) is 6.50 Å.

## III. Comparison of optical absorbances from TD-a*GW* and *GW*-BSE

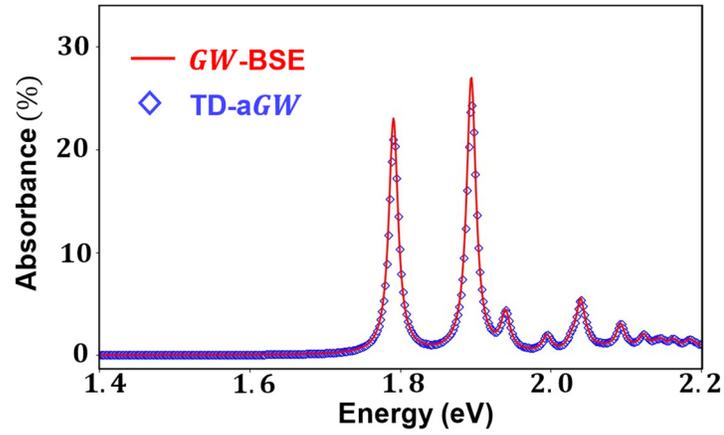

Fig. S2: Linear optical absorbance from *GW*-BSE (red line) and TD-a*GW* (blue dimond).

As discussed in the main text and Section 1C of this Supplemental Material, the many-body excitonic effects (electron-hole interactions) in the TD-a*GW* is equivalent to that of *GW*-BSE in the linear optical response. As a result, they give identical linear absorbance spectra as shown in Fig. S2.

## IV. Hole-hole coherence dynamics

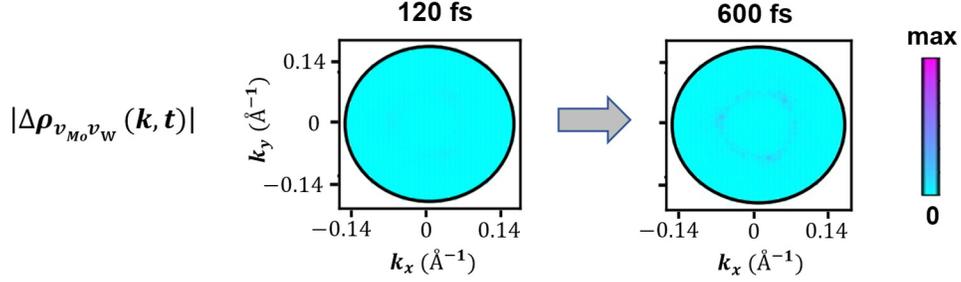

Fig. S3: $k$-distributions of hole-hole coherence dynamics: $\Delta\rho_{v_{Mo}v_W}(k,t)$ at $t = 120$ fs and 600 fs. The magnitude is plotted, and the same color bar is used as in Fig. 2(d) of the main text.

In the main text, the dynamics of the interlayer and intralayer electron-hole coherences have been investigated. Here we will discuss the coherence between holes in different layers which is characterized by $\Delta\rho_{v_{Mo}v_W}(k,t)$. Comparing Fig. S3 and Fig. 2(d) of the main text, we find that the magnitude and dynamics of hole-hole coherence are significantly smaller than those of electron-hole coherences. Therefore, the direct valence-valence band transition from one layer to anther is negligible for this system, and single-particle hole transitions without phonon assistance between valence bands are in fact not a significant contribution to the dynamics of interest.